\documentclass[%
 aip,
 amsmath,amssymb,
preprint,%
]{revtex4-1}

\usepackage{graphicx}
\usepackage{dcolumn}
\usepackage{bm}
\usepackage[utf8]{inputenc}
\usepackage[T1]{fontenc}
\usepackage{mathptmx}

\begin{document}


\title{Magnetized Gas Collimation of Interstellar Outflow Scaled by Laser-produced Plasma}

\author{Tao Tao}
\affiliation{Department of Engineering and Applied Physics, University of science and technology of China, Hefei,Anhui, 230026, China}

\author{Guangyue Hu}
\email[e-mail:]{gyhu@ustc.edu.cn}
\affiliation{Department of Engineering and Applied Physics, University of science and technology of China, Hefei,Anhui, 230026, China}

\author{Ruxin Li}
\affiliation{State Key Laboratory of High Field Laser Physics, Shanghai Institute of Optics and Fine Mechanics, Chinese Academy of Sciences, Shanghai 201800, China}
\affiliation{Collaborative Innovation Center of IFSA, Shanghai Jiao Tong University, Shanghai 200240, China}

\author{Zhizhan Xu}
\affiliation{State Key Laboratory of High Field Laser Physics, Shanghai Institute of Optics and Fine Mechanics, Chinese Academy of Sciences, Shanghai 201800, China}
\affiliation{Collaborative Innovation Center of IFSA, Shanghai Jiao Tong University, Shanghai 200240, China}

\author{Jian Zheng}
\affiliation{Department of Engineering and Applied Physics, University of science and technology of China, Hefei,Anhui, 230026, China}
\affiliation{Collaborative Innovation Center of IFSA, Shanghai Jiao Tong University, Shanghai 200240, China}

\date{\today}

\begin{abstract}
Widely accepted modeling of Young stellar objects/planetary nebula outflow anisotropies usually involve wind-wind interactions and magnetic collimation, but due to observational constraints, detailed structures of wind and magnetic fields inside collimation region remain undetermined. We numerically investigated its laboratory counterpart, based on poloidal field collimation in magnetocentrifugal launching model. Our analog consist of fast wind: Aluminum plasma generated by $I\sim3.2\times10^{13}\ W/cm^{2}$ ns laser, and magnetized ambient: molecular Helium of density $\rho\sim1\times10^{-8}$-$2.56\times10^{-6} g/cm^{3}$ and $B\sim5-60$ Tesla embedded field. Elevating magnetic field strength or decreasing gas density can alter expansion morphology, from sphere to prolonged cavity and ultimately to collimated jet. In theoretical analysis, we also found the transition of ambient response from anisotropic to isotropic can be attributed to the developing of strong MHD discontinuity. Outflow patterns like collimation levels can now be quantitatively predicted based on the knowledge of its surroundings, more specifically, by a set of external Mach numbers. We conclude that such mixed gas and magnetic field dynamics are consistent with astronomical observations of protostars and planetary nebulae in certain evolution stages. We provide a scalable framework allowing fitting of flow-field structures in astronomical unresolved regions by assuming their possible geometries on a repeatable laboratory platform. 
\end{abstract}

\maketitle

\section{Introduction}
Departure of central outflow from spherical expansion in young stellar object(YSO)\cite{reipurth1997,reipurth2001} and planetary nebula(PN)\cite{soker1994,balick2002} are ubiquitous, though in terms of collimation level, they varies greatly from near spherical to bipolar shape. One of the central problems is the modeling of astronomical unresolved collimation area, it should be able to explain large-scale observations while stay consistent with the central star boundary conditions. One candidate is the pure hydrodynamics(HD) wind-wind interaction model, that describes how a slow moving YSO envelop\cite{arce2002}, or disk wind, collimates the fast initially wide-angle central outflow into bipolar jets\cite{Bally1983,Lada1985,Bachiller1996,frank2014}. Evidence like shocks\cite{Dopita1982,Mundt1983,Reipurth1986} can be seen in the colliding winds that have different origins and properties. For PNs, such interacting stellar winds model(ISW)\cite{Kwok1978,Kahn1985,Balick1987} also exists, consist of a central isotropic fast expanding outflow and a slower but denser envelop\cite{bujarrabal2001} ejected in previous star evolution stages. Another candidate is the magnetohydrodynamics(MHD) model, it propose shaping by global coherent stellar magnetic fields\cite{carrasco2010} \cite{kastner2012}. Magnetic field serve as a key component in YSOs' launching mechanism\cite{ferreira2006,stute2010} and can continue its influence in the process of collimation\cite{Bally2016}. PNs' complex symmetry pattern\cite{balick2002} also show sign of magnetic confinement, yet the exact field topology and strength can not be fully determined\cite{hartigan2007}\cite{soker2006,sabin2007}. The actual collimation is a non-equilibrium process. Even for a certain object at a certain time, gas ram, thermal pressure and magnetic pressure, any one of them can be the dominate factor of evolution in different spatial locations. Such complexity can only be fully explained by the coupled flow-magnetic dynamics by integrated simulation or experiment. 

Laboratory astrophysics\cite{remington1999,drake1999,takabe2001} experiments on collimation of laser plasma presents good analogy to stellar outflows, exploring key physical candidates by mimicking winds and embedded field in scaled experiments and simulations: high-Z plasma jets introduced by radiative cooling effect\cite{lebedev2005}; colliding entrainment between multiple plasma species generated by tuned ablation pattern and shaped target\cite{suzuki2012,Yurchak2014}; and magnetic confinement of laser-generated plasma using externally applied field\cite{Ciardi2013,Albertazzi2014}. These efforts demonstrate a variety of mechanisms that can lead to collimation. A series of dimensionless quantities that mark various non-ideal processes, such as radiation, magnetic diffusion, etc., need to be in the same range as astronomical facts when it comes to determine whether the laboratory scaling stands on solid bases.

In this article, we seek further integration of wind and field confinement by numerically investigated a scaled laser experiment, it consists of laser produced plasma driver and a magnetized molecular ambient, emphasizing the joint effect of background gas and applied magnetic field. An increase in background magnetization level alters the expansion morphology, from spherical bubble to highly collimated jets. By modeling ambient isotropic to anisotropic response character, criterion for collimation level and production of jets is derived. Various astronomical outflow patterns can be identified and explained on this scalable mixed HD and MHD framework, by assuming possible conditions of their surrounding mediums.

This article is organized as follows: Sec. II presents the detail of numerical tool, Sec. III presents the setup geometry and laser ablation scaling, Sec. IV presents the fully developed expanding plasma and its dynamics in different ambient field and density combinations, Sec. V presents quantitative analysis of how expansion morphology change due to the ambient, and Sec. VI presents how these laboratory scale results compare to astronomical observations.

\section{Simulation Methods}
Numerically, we employ the single fluid resistive MHD, three dimensional Eulerian code FLASH\cite{fryxell2000} with tabulated EoS and opacity table IONMIX\cite{macfarlane1989}. Speaking of its capabilities, FLASH can handle laser ablation, magnetic field dynamics and non-ideal material properties in a single consistent run, it is robust against shocks even with high precision 3rd order format, ideal for our task where no initial equilibrium can be found. Energy source term in MHD set is laser inverse bremsstrahlung absorption, here realistic ray trajectory have been considered. Energy deposited in electrons then get relaxed among ions and multigroup radiation species, electron number densities needed here are provided by querying tabulated ionization table. Diffusion related coefficients like heat conductivity, viscosity and magnetic diffusitivity are derived runtime by LeeMore collision model\cite{lee1984}. The radiation MHD equations we evolve are:
\begin{equation}
\dfrac {\partial \rho }{\partial t}+\nabla \cdot \left( \rho \bm{v}\right) =0
\end{equation}
\begin{equation}
\dfrac {\partial \rho \bm{v}}{\partial t}+\nabla \cdot \left( \rho \bm{vv}-\bm{BB}\right) +\nabla P_{tot}=\nabla \cdot \tau
\end{equation}
\begin{equation}
\begin{split}
\dfrac {\partial \rho E}{\partial t}+\nabla \cdot \left( \bm{v}\left( \rho E+P_{tot}\right) -\bm{B}\left( \bm{v}\cdot \bm{B}\right) \right) = \\
\nabla \cdot \left( \bm{v}\cdot \tau +\sigma \nabla T\right)+ \nabla \cdot \left( \bm{B}\times\left( \eta \nabla \times \bm{B}\right) \right) + \\
Q_{las}-\nabla \cdot \left( q_{ele}+q_{rad}\right)
\end{split}
\end{equation}
\begin{equation}
\dfrac {\partial \bm{B}}{\partial t}+\nabla \cdot \left( \bm{v}\bm{B}-\bm{B}\bm{v}\right) =-\nabla \times \left( \eta \times \bm{B}\right)
\end{equation}
where
\begin{equation}
P_{tot}=\dfrac {B^{2}}{2}+P_{ion}+P_{ele}+P_{rad}
\end{equation}
\begin{equation}
E=e_{ion}+e_{de}+e_{rad}+\dfrac {1}{2}\bm{v}\cdot \bm{v}+\dfrac {1}{2}\dfrac {B^{2}}{\rho }
\end{equation}
\begin{equation}
\tau =\mu \left( \left( \nabla \bm{v}\right) +\left( \nabla \bm{v}\right) ^{T}-\dfrac {2}{3}\left( \nabla \cdot \bm{v}\right) \bm{I}\right)
\end{equation}
are total pressure $P_{tot}$, specific total energy $E$ and viscous stress tensor $\tau$. $T$ is temperature, $\epsilon$ is the specific internal energy, $\bm{B}$ is the magnetic field, $Q_{las}$ is laser energy source, $q$ is heat flux, $\mu$ is the coefficient of dynamic viscosity, $\sigma$ is the heat conductivity, and $\eta$ is the resistivity.

Simulation boundary condition are selected to be gradient free or reflecting, depend on whether to model coil as a conductive wall, though they show minimal differences because plasma dimension is significant smaller than that of the coil. 3D Cartesian runs have $\Delta x=50\ \mu m$ spatial resolution, 2D cylindrical runs with higher resolution $\Delta x=20\ \mu m$ are also employed to fill the gaps in parametric scan. 

\section{Model Setup and Ablation Properties}
\begin{figure}[bht]
	\centerline{\includegraphics[width=8cm]{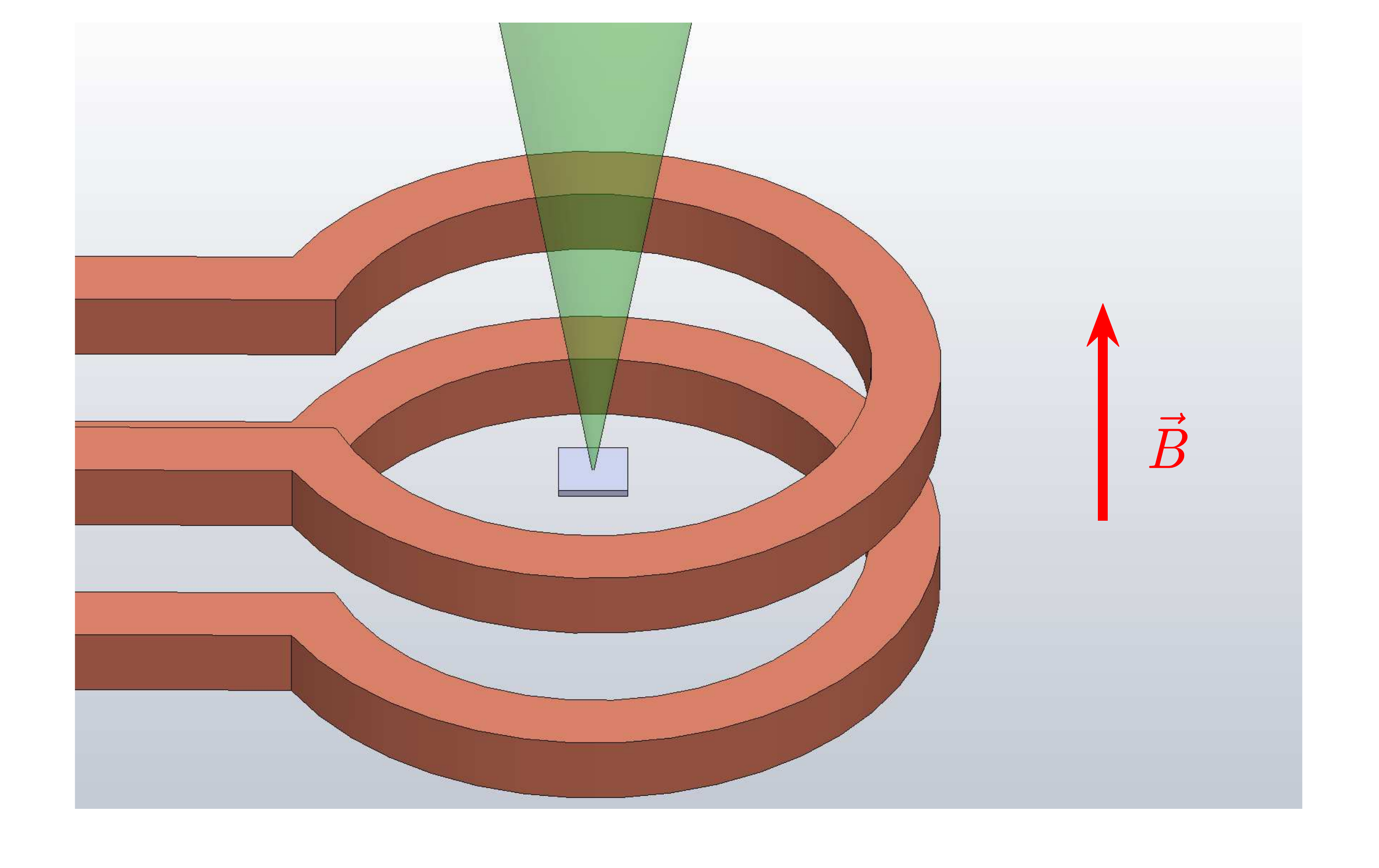}}
	\caption{\label{setup} Targeted experiment geometry setup, illustrating placement of laser, coils and Aluminum planar target. Neutral ambient is filled inside the chamber and not explicit shown here. 
	}
\end{figure}

\begin{figure*}[bth]
	\centering 
	\includegraphics[width=18cm]{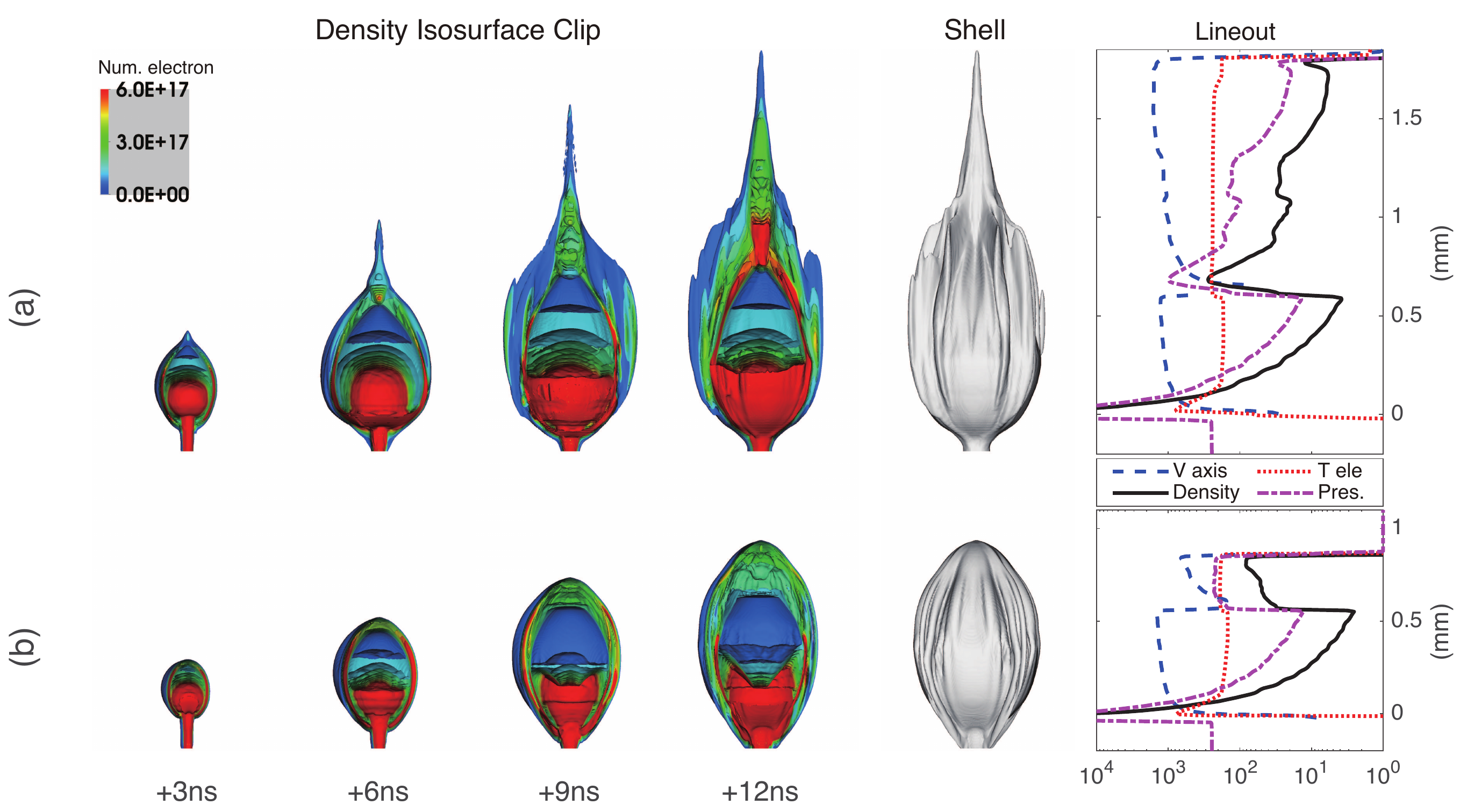}
	\caption{\label{3d} 3D density isosurface map of Aluminum plasma evolution and line-out data along symmetry axis; Row (a) with $1.0\times10^{-8} g/cm^{3}(n_{ele}=3.0\times10^{15})$ ambient, row (b) with $1.28\times10^{-6}\ g/cm^{3}(n_{ele}=3.9\times10^{17})$ ambient. $P_{l}=10^{10}\ Watt$ Laser incident from the top, $20$ Tesla magnetic field is uniformly pointing upwards at $t=0$. Half clip is applied to reveal internal density structure in 4 different time frames, complete shell surface and line-out are results with $+12$ ns delay from the rising edge of laser. Units of line-out quantities are: velocity in $km/s$, electron temperature in $eV$, pressure in $GPa$ and density in $10^{16}/cm^{3}$.
	}
\end{figure*} 

According to astronomy backgrounds, the scaled laboratory counterpart of stellar outflow should include three components: laser-produced aluminum plasma act as the fast wind, prefilled stationary helium act as the slow ambient, and a global external B field. Laser, magnetic field and the target are placed as shown in Fig.~\ref{setup}, the simulation box only include the inner coil surfaces as numerical boundaries. Direction of initial magnetic field, laser incident direction and the symmetry axis of ablated plasma are parallel to each other, this field configuration often referred as "poloidal", a counterpart of the astronomical magnetocentrifugal field in the collimation region. Toroidal field component is possible by using additional coils, but since we do not take launching and self-pinching process into consider, poloidal component only is a simple and sufficient approximation. 

Nanosecond heater laser wavelength is $\lambda_{l}=1064\ nm$, it keeps a constant power of $P_{l}=1.0\times10^{10}\ Watt$ throughout the entire process. Laser spot radius is $r_{l}=100\ \mu m$, a power density of $I_{l}=3.2\times10^{13}\ W/cm^{2}$ is achieved. Plasma dynamic time scale is several $10s$ of ns, and $\sim 1\ cm$ of spatial scale. We hope to mimic a quasi-stable outflow source by using this long pulse pedestal, worth to mention even turn off the heater won't stop the out expanding source immediately, cause higher density region inside the target below ablation front can get heated by laser impact shockwave or electron conduction, creating a decaying source lasting up to $100\ ns$. A detail tailored laser time profile have the potential of mimicking behavior of a variable astronomical source, but we will leave this topic for another time. 

A solid state Aluminum target with $\Phi=300\ \mu m$ surface diameter is small enough not to drag magnetic field through induction, in real practice it can be achieved by hitting the tip of a wire target. Hot plasma is driven out by pressure gradient near the laser spot, maximum outgoing speed can be estimated by a one sided rarefaction approximation: isothermal sound speed of Al plasma upon direct ablation scale as $c_{0}=8.64\times10^{7}I^{1/3}\lambda^{2/3}\ cm/s$\cite{atzeni2004}, maximum rarefaction front speed can reach $2c_{0}(\gamma)^{1/2}/(\gamma-1)=3.55c_{0}$\cite{fabbro1985} without any external confinement. Initial plasma opening angle can be slightly converged towards target normal from $\sim180^{\circ}$ to $\sim160^{\circ}$, due to half-space planar target geometry and possible laser impact crater, but still far from collimation. Relatively low atomic number of Aluminum exclude the possibility of spontaneous collimation by cooling on $<100\ ns$ timescale, assured in later analysis by a large Peclet number, so any reduce in outflow angle can only be the result of gas ramming or field confinement.

Pre-filled background were chosen to be molecular state Helium of variable density $\rho=1\times10^{-8}$-$2.56\times10^{-6}\ g/cm^{-3}$(electron number density $n_{ele}=3\times10^{15}$-$7.7\times10^{17}$), with variable $B=5$-$60$ Tesla magnetic field. 
Reasons of this specific parameter space are: One, such field is close to the highest possible strength from current capacitor based pulsed generator without implosion compression; Two, molecular Helium have high enough ionization threshold not to be disturbed by pulse coil induced vortex electric field prior to primary heating laser; Three, and the most important, those combinations can cover region of super-Alfvenic to sub-Alfvenic transition.

\section{Expansion Morphology}

\begin{figure*}[hbt]
	\includegraphics[width=18cm]{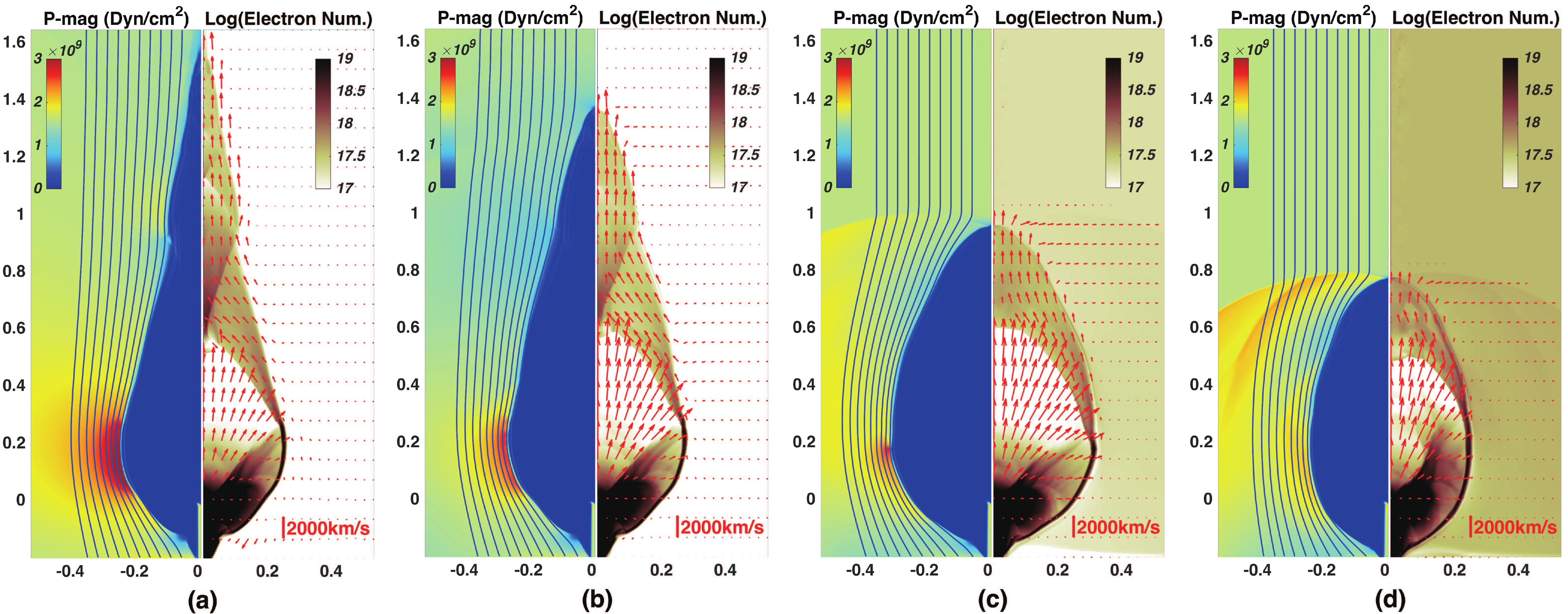}
	\caption{\label{denspres} Pair of pseudo-color map of magnetic pressure and electron number density at $t=12ns$, for 4 different ambient Helium densities: $(a)1.0\times10^{-8}\ g/cm^{3}(n_{ele}=3.0\times10^{15})\ (b)1.6\times10^{-7}\ g/cm^{3}(n_{ele}=4.8\times10^{16})\  (c)6.4\times10^{-6}(n_{ele}=1.9\times10^{17})\ (d)1.28\times10^{-6}\ g/cm^{3}(n_{ele}=3.9\times10^{17})$; Magnetic field lines and velocity vectors are overlapped on all plot pairs.
	}	
\end{figure*}

Morphology of the expanding plasma can be altered by surrounding gas envelop properties in several different stages. First we study the time evolution behavior using two representatives in Fig.~\ref{3d}, Fig.~\ref{3d}(a) with near vacuum background and Fig.~\ref{3d}(b) with ambient whose density are comparable to the outflow, field strength is $20$ Tesla for both cases. Since MHD code does not allow zero density for numerical reasons, we can only approach it by iteratively decreasing ambient until the result morphology converges. Plasma thermal and ram pressure always dominates at the beginning of ablation before $3$ns because deposited energy are concentrated near the spot, pressure and density along the central axis show exponential decaying profile consistent with previous mentioned rarefaction model. It is worth to note plasma bubble is very thin and laser can easily penetrate it, so whatever inside the laser tunnel above the critical surface does not get heated by laser deposition directly. As spatial scale increases after 9ns, the outflow energy density decreases inversely proportional to $L^{3}$ and approaching quasi-steady state, clear shell walls start to take form in transverse direction on a pressure balanced interface. It can be confirmed from velocity line-out that the profile of supersonic flows inside the shell are similar between (a) and (b): $\sim 600eV$ hot spot with isothermal sound $c_{0}=290\ km/s$ accelerates Aluminum plasma outwards up to $1000\ km/s$ in the transverse direction, velocity along the axis can further reach $5-6c_{0}\sim1500km/s$ in a funnel shape passage, electron density spreads thinner down to $2\times 10^{16}\ /cm^{3}$ along the way. The most distinguished feature between the two runs comes from whether there is an extension structure outside the shell. Fig.~\ref{3d}(a) possess jet on top of the elliptical cavity, rooted on a distinct cone-shaped shock transfer surface\cite{canto1988}, while (b) is a closed cavity, here conical shock replaced by a pair of in-and-outward facing bow-shocks, density accumulated in between the shock increase more than one order in magnitude. Additional features demonstrated on shells are mainly instabilities: in both cases, Kelvin-Helmholtz(K-H) instability on velocity shear boundaries are suppressed due to the magnetic tension stabilizing effect\cite{keppens1999}, reduce its chance of efficiently disturbing the jet. Yet Rayleigh–Taylor(RT) filaments with a longitudinal pattern grows all the way to non-linear regime, with wave vector perpendicular to the field direction. It's fundamentally a type of interchange mode between inside constantly ramming outflow and outside compressed/bent magnetic field lines.

Increasing background density within a certain range can effectively eliminates the poloidal extended jet but preserve transverse dimension of the shell, like illustrated in Fig.~\ref{denspres} with two dimensional field and flow topology slice. On transverse direction, the interface type is tangential discontinuity, separating inner diamagnetic cavity from the magnetized ambient. All 4 Fig.~\ref{denspres}(a)-(d) have nearly the same shell radius $r=0.25\ cm$, because shell size are solely determined by the balancing of deposited laser heat $E_{l}=P_{l}t$ and expelled magnetic field energy $r^{3}B^{2}/4\pi$, so we can expect $r\propto E_{l}^{1/3}/B_{0}^{2/3}$. At the same time maximum Plasma beta in the perturbed ambient are $\beta\sim0.0015-0.05<<1$ for all cases, as long as thermal pressure remain a relative small quantity, transverse diameter will always subject to above scaling, and flow structure will ultimately converge to this quasi-stable velocity shear with small interface speed. However in poloidal direction, whether it is possible to maintain a stable high speed jet do not solely rely on plasma beta. In Fig.~\ref{denspres}(a) central plasma outflow get refracted on the wall and cancel each others' transverse velocity to form collimated jet. Fig.~\ref{denspres}(c)-(d) show how that jet get stagnated by a clearly visible shock front which can be seen sweeping through the background. In Fig.~\ref{denspres}(a) any field disturbance by compression get quickly evened and travels further in Alfven speed $v_{a}=\sqrt{B^{2}/\mu_{0}\rho}>5000\ km/s$, so field will not piled up on the expanding shell. Fig.~\ref{denspres}(b) has a lower Alfven speed $v_{a}=1400\ km/s$, meaning characteristic plasma driver speed $v_{d}\sim 1500 km/s$ has surpassed the Alfven point. From now magnetized background gradually loses the anisotropic response character, its dynamic behavior changes from flow following magnetic field to field following the flow.

\section{Stagnation by MHD shock}
Conical shock collimation structure does not disappear immediately after the crossing of Alfven point, actually, it takes an ambient density about 4 times higher to fully suppress the jet. It is necessary to quantitatively identify the thresholds for poloidal jet generation and also when confinement happen in transverse direction, and where the morphological boundaries of HD and MHD are. Parameter scan of variable field-strength is added along with previous variable density cases, results were successfully explained by MHD shock model.

\begin{figure}[thb]
	\centerline{\includegraphics[width=8cm]{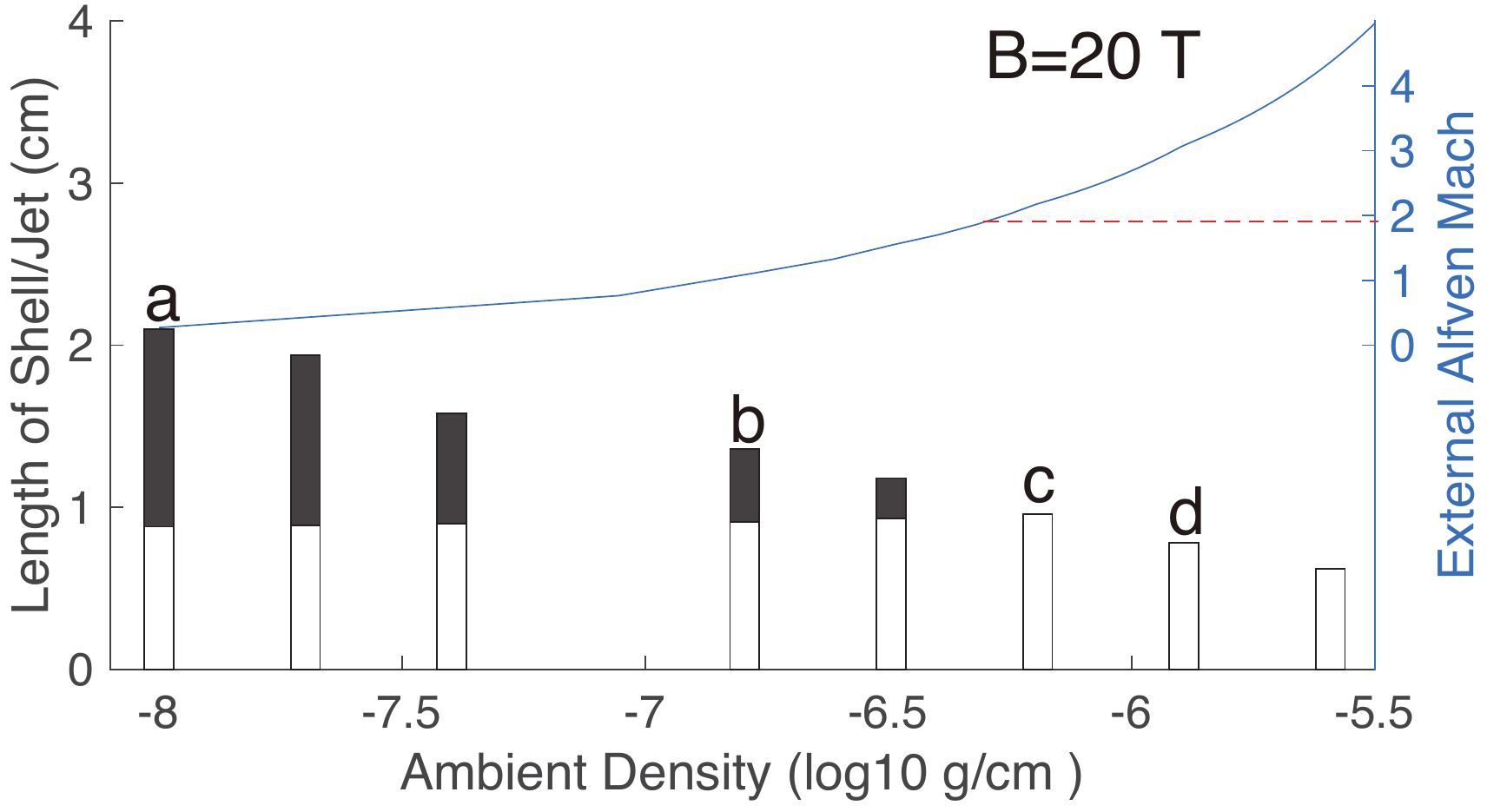}}
	\caption{\label{histogram} Shell/jet(if present) length on axis represented by white and black bars respectively, snapshot at 12ns when shell reach its maximum size. From left to right are fixed 20 Tesla field with ramping up $(1, 2, 4, 16, 32, 64, 128, 256)\times10^{-8}\ g/cm^{3}$$(n_{e}=(3.0, 6.0, 12, 48, 96, 193, 385, 771)\times10^{15}\ /cm^{3})$ ambient density, Right axis shows the shell Alfven Mach number $v_{d}/v_{a}$. Annotation $a-d$ correspond to plots in Fig.~\ref{denspres}
	}
\end{figure}

First parametric scan in Fig.~\ref{histogram} use fixed field strength and varying ambient. Shell length is calculated from target surface to peak point of the two branches, from there to the uppermost perturbation front is considered as the jet length. Jet length decreases linearly with exponential growth of background density, accompanied by strong MHD perturbation front develop tangent to the shell, with magnetic field thread through. We employ external Alfven Mach number $M_{a}$ here, which is the maximum plasma driver speed $v_{d}$ versus the ambient Alfven speed. When jet is fully suppressed, Alfven speed of undisturbed ambient $v_{a}\sim700\ km/s$, gives $M_{a}$ a value $\sim 2$.

\begin{figure}[htb]
	\centerline{\includegraphics[width=8cm]{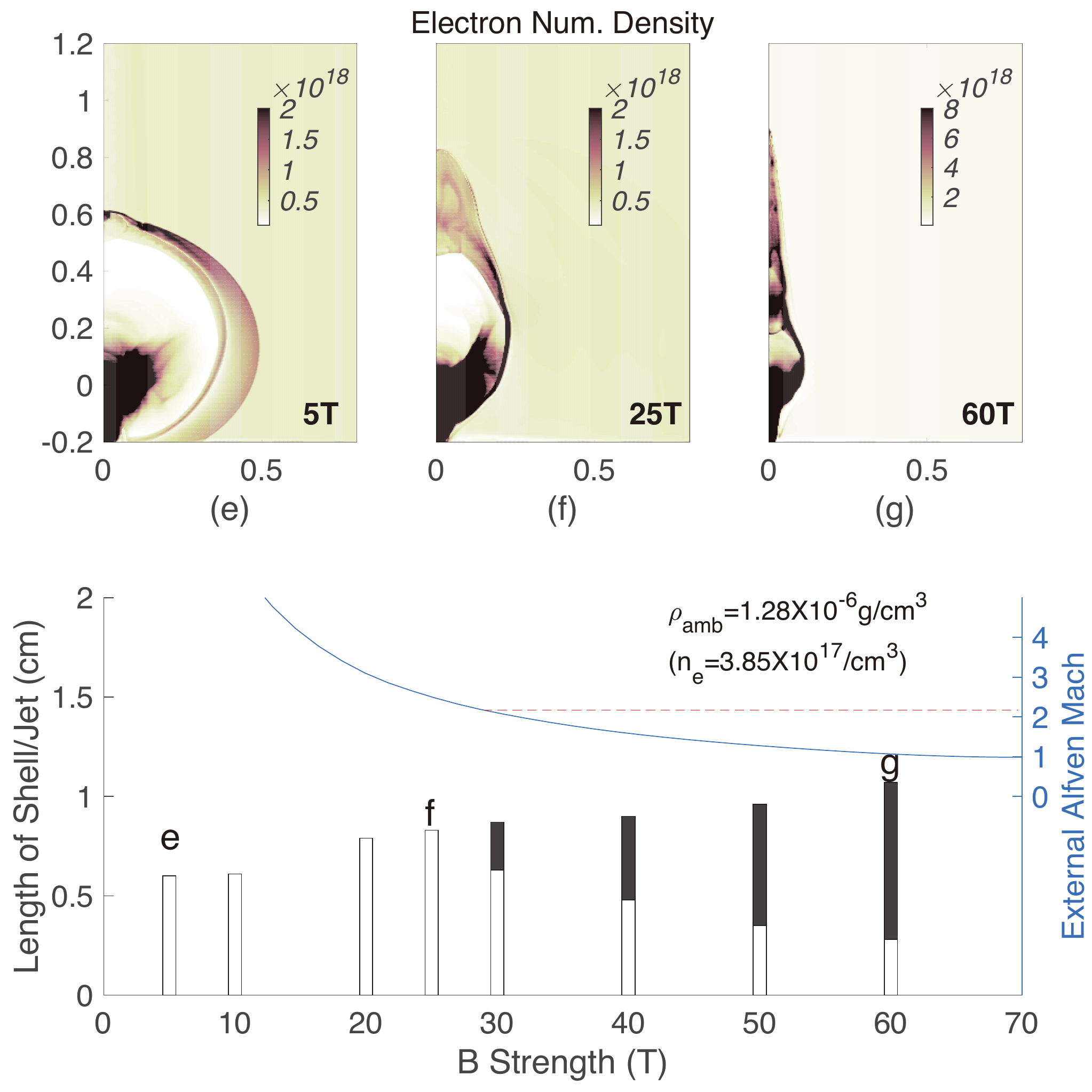}}
	\caption{\label{histogram_dens} 
		Histogram showing Shell/jet length in white and black bars like in previous fig., but for fixed 1e-5 ambient density and ramping up (5, 10, 20, 25, 30, 40, 50, 60) Tesla B field, with Alfven Mach number overlap on the right. 3 corresponding density illustrations $e-g$ are shown above.
	}	
\end{figure}

Another scan use fixed $1.28\times10^{-6}\ g/cm^{3}$ ambient density and changing field strength in Fig.~\ref{histogram_dens} .Thin shell expanding isotropically in a gas dominated dynamics, initially $5\ T$ field indeed get amplified along the compression of gas but do not confine the expansion; $25\ T$ is enough for prolonged shell to show up yet without conical shock structure; a highly collimated jet is finally retrieved up to $60\ T$, though in a much smaller total scale than Fig.~\ref{denspres}(a). Critical point of producing jet show the same external $M_{a}\ \sim 2$.

\begin{figure}[htb]
	\centerline{\includegraphics[width=8cm]{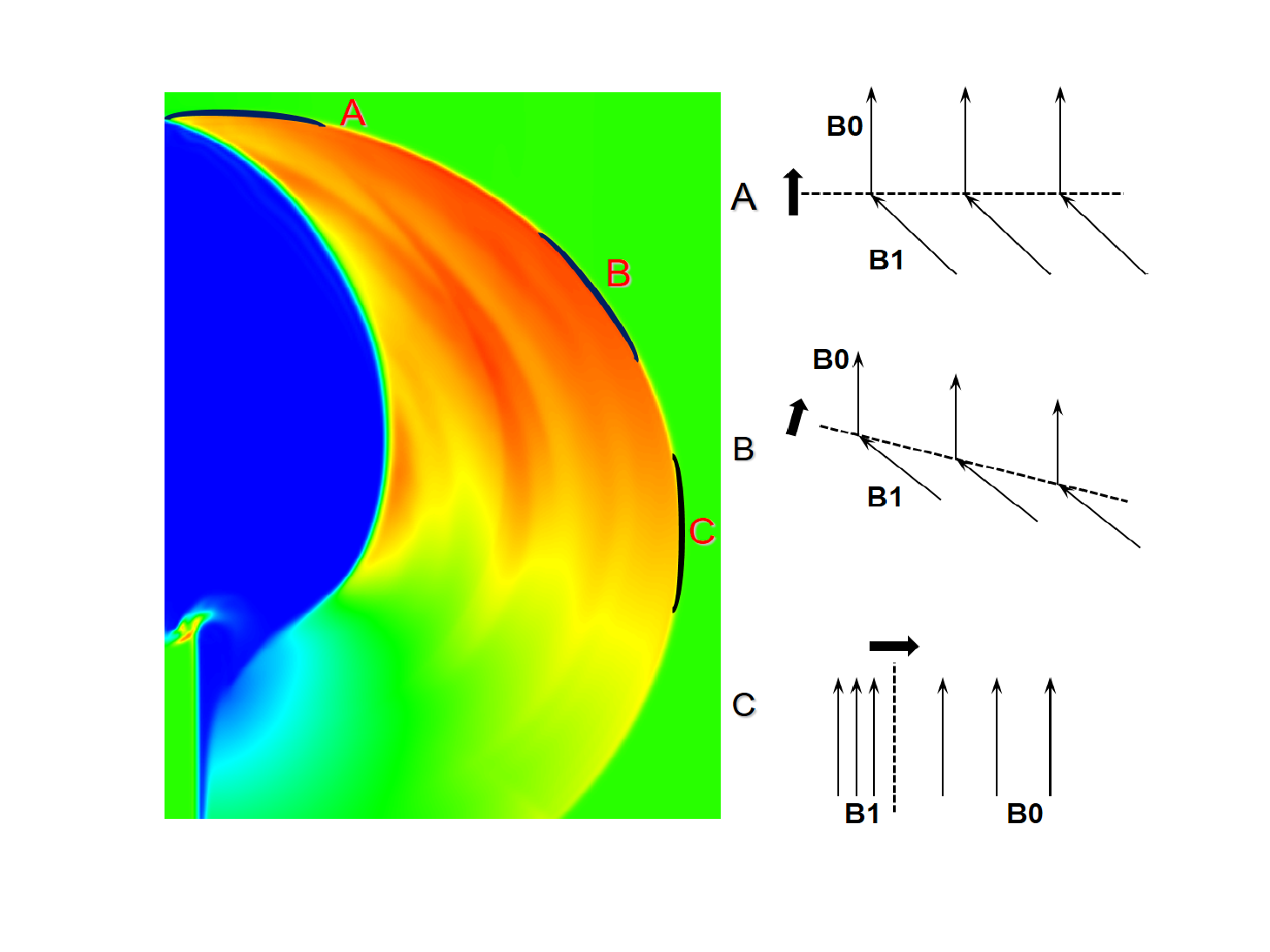}}
	\caption{\label{3shock} 
		Different geometries of the magnetic field relative to shock surface(dotted line): A, switch-on shock; B, oblique-shock; C, perpendicular-shock.
	}	
\end{figure}

The expansion front of outflow is similar to a piston. A strong supersonic piston compression amplifies the field adjacent to the surface and form a subsonic region, this subsonic region is separated from the further undisturbed medium by a shock wave. Compared to a case in neutral gas with the same piston driving intensity, magnetic field always increases the restoring force of the medium, providing higher shock velocity and larger subsonic region width, magnetic field anisotropy response appears partially in this subsonic region. When continuously increase the compression intensity, shock velocity and the subsonic region width values will converge to cases where magnetic field is absent. It is at this time jet and the anisotropy are completely eliminated. Predicting the magnetized shock speed demand inserting additional  magnetic field energy $\bm{G}$ and momentum flux $\bm{\Pi}$: 
\begin{equation}
\bm{G_{b}}=\left( B^{2}\bm{v}-\left( \bm{B}\cdot \bm{v}\right) \bm{B}\right)/{4\pi }
\end{equation}
\begin{equation}
\bm{\Pi_{b}}=\left( {B^{2}}\bm{I}/2-\bm{BB}\right)/{4\pi }
\end{equation}
into HD Rankine-Hugoniot relations\cite{landau1984}, along with magnetic, electric field boundary conditions and continuity equation:
\begin{equation}
B_{0\perp }=B_{1\perp }
\end{equation}
\begin{equation}
\left( \bm{v_{0}\times B_{0}}\right) _{\parallel}=\left( \bm{v_{1}\times B_{1}}\right) _{\parallel}
\end{equation}
\begin{equation}
\rho v_{0\perp }=\rho v_{1\perp }
\end{equation}
where subscript $0$ and $1$ represent up-and-downstream quantities, $\perp$ and $\parallel$ represents direction relative to the discontinuity surface. The most common scenario in our simulation is the shock plane has an angle with respect to the undisturbed magnetic field as illustrated in Fig.~\ref{3shock}(b), downstream amplified magnetic field is bent toward the interface. When shock plane is parallel to the field lines as Fig.~\ref{3shock}(c), its speed $v_{s}\geq\sqrt{c_{s}^{2}+v_{a}^{2}}$, indicating it will converts to fast magnetoacoustic wave as shock intensity diminished. 

A special case is the "switch-on" shock\cite{liberman1978} shown in Fig.~\ref{3shock}(a), named so because azimuthal component of B experience zero to non-zero switching. For poloidal field configuration discuss in this article, switch-on shock's decay to the HD characteristic also means the full suppression of jet. Stable existence of switch-on shock requires two conditions: flow deceleration $v_{1}<v_{0}$ and field strength increasing $B_{1}>B_{0}$, Substitute into MHD R-H relations produce $1<M_{a0}^{2}<4M_{0}^{2}/(M_{0}^{2}+3)$, here $M$ is the HD Mach number. These inequalities indicate distinguished morphology character exist under following conditions:\\
1.$v_{a}\gg c_{s}$ and $v_{d}<v_{a}$, field dominated sub-Alfvenic expansion with no shock, full jet and elliptical shell, Fig.~\ref{denspres}(a).\\
2.$v_{a}\gg c_{s}$ and $v_{a}<v_{d}<2v_{a}$, field dominated MHD shock in cross Alfvenic transition, with decaying jet and elliptical shell, Fig.~\ref{denspres}(b).\\
3.$v_{a}\gg c_{s}$ and $v_{d}>2v_{a}$, field dominated but HD like super-Alfvenic expansion, without jet only elliptical shell, Fig.~\ref{denspres}(d).\\
4.$v_{a}\leq c_{s}$, inequalities can never be satisfied, pure HD gas behavior always dominates, without any form of collimation at any time, Fig.~\ref{histogram_dens}(e).

\begin{figure}[htb]
	\centering
	\includegraphics[width=8.0cm]{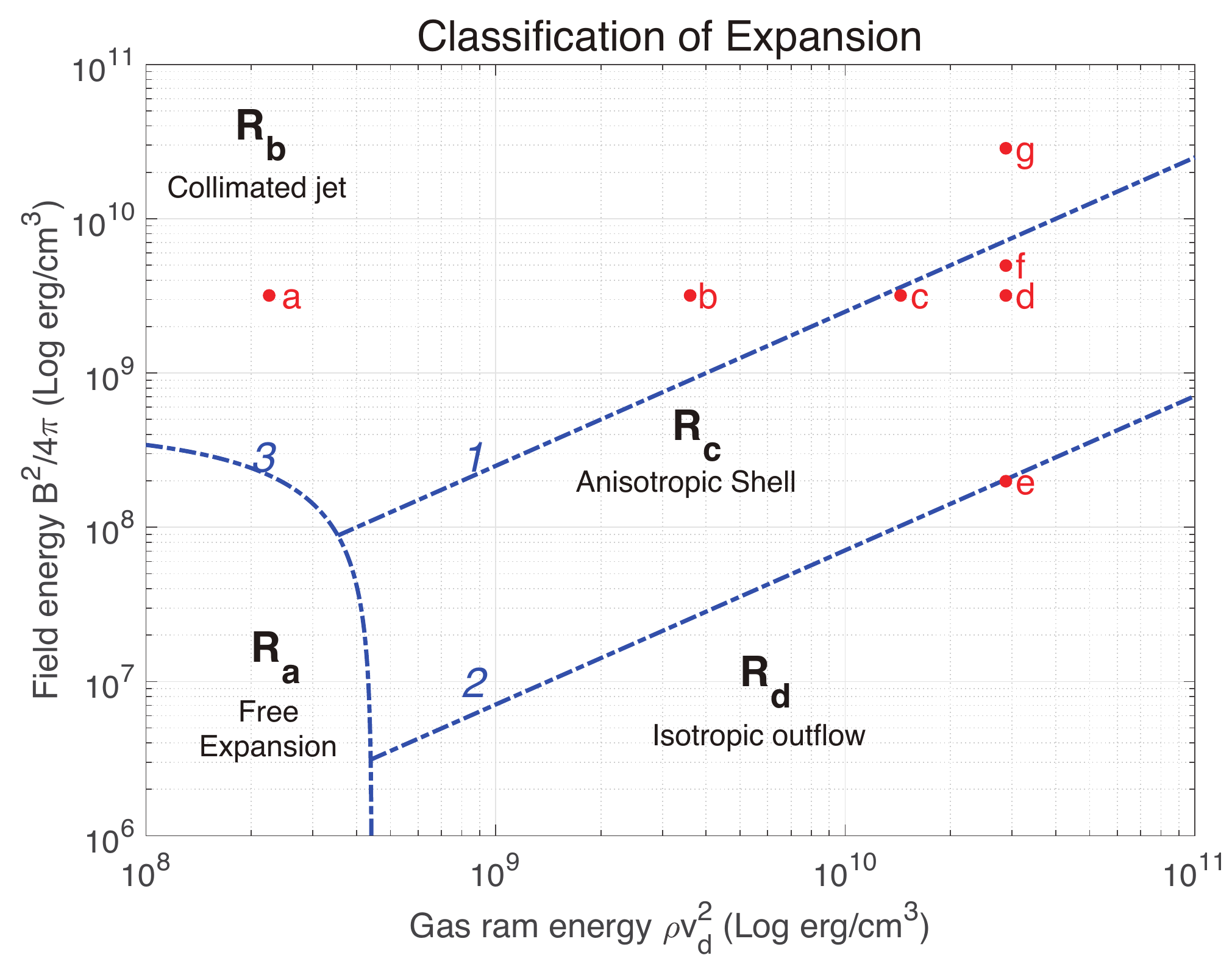}
	\caption{\label{phase} Plasma expansion classification map for various ambient conditions. Two axis represents magnetic field and gas ram energy density; $R_{a}-R_{d}$ represent 4 sections with distinguished plasma morphology and separate by dashed lines: 1 is jet criterion $v_{d} \geq 2v_{a}$; 2 is anisotropic shell criterion $v_{a} \leq c_{s}$; 3 is free expansion criterion $E_{ram}+E_{B} \ll E_{out}$, $E_{out}$ is the energy density of outflow. Necessary parameters are used, $v_{d}~1500 km/s$, $T\sim500eV$ and estimated $E_{out}=0.1\times P_{l}\tau/l^{3}$ as a fraction of deposited laser energy and decrease with scale length, marker $a-g$ correspond to the 6 points previously shown as plots in Fig.~\ref{denspres} and Fig.~\ref{histogram_dens}.
	}
\end{figure}

The outflow collimation level can now be predicted quantitatively from known source intensity, field strength and ambient density. Or in turn we can infer source character and ambient situation from observed outflow pattern. To show its consistency, Fig.~\ref{phase} summarize the morphology criterion and distinguished outflow topology using a two-dimensional map whose two axes are magnetic field energy density $B^{2}/4\pi$ versus gas ram energy density $\rho v^{2}$, rewrite $v_{d}=2v_{a}$ and $v_{a}=c_{s}$ in the form $B^{2}/\left( 4\pi \rho V^{2}_{d}\right) =1/4$ and $B^{2}/\left( 4\pi \rho V^{2}_{d}\right) =\gamma T$, thus these two conditions correspond to two lines with a fixed slope on the energy density map. Cases a-g appeared in previous sections all fall into their expected classification regions. In astrophysics context, supernova explosion correspond to free expansion region $R_{a}$; $R_{b}$ accommodates magnetized high level collimation with jet in many YSOs; ; $R_{c}$ is the shell asymmetry only region, seen in many PNs; $R_{d}$ includes pure HD, symmetry outflows. In general, instance points further away from the origin will have smaller spatial and temporal scale.

\section{Connection with Astro collimated Outflow}

\begin{table}[htb]
	\caption{\label{tab:table1}Key parameters for plasma collimation in magnetized ambient, subsript $d$ stands for drive matter, $s$ for shell, $a$ for ambient, $r$ for radiation.}
	\begin{ruledtabular}
		\renewcommand\arraystretch{1.0}
		\begin{tabular}{lllll}
			\multicolumn{2}{l}{Parameter}&Laboratory&YSO&PN\\ 
			\hline \\
			\multicolumn{2}{l}{Spatial $L(cm)$}&$<1$&$>10^{17}$&$>10^{18}$\\
			\multicolumn{2}{l}{Velocity $v_{d}(cm/s)$}&$1.5\times10^{8}$&$<10^{8}$&$<10^{8}$\\	
			\multicolumn{2}{l}{Field strength $B_{a}$}&$50-600\;kG$&$<100\;mG$&$<10\;mG$\\
			\multicolumn{2}{l}{Outflow Density $n_{d}$}&$5\times10^{16}-10^{18}$&$10^{1}-10^{5}$&$>10^{5}$\\
			\multicolumn{2}{l}{Time Scale $\tau=L/v_{d}$}&$10\; ns$&$>10^3\; yr$&$>10^4\; yr$\\	
			\multicolumn{2}{l}{Ext. $M$, $v_{d}/c_{s}$}&$>10$&$>10$&$>10$\\
			\multicolumn{2}{l}{Ext. $M_{a}$, $v_{d}/v_{a}$}&$0.1$-$10$&$>0.1$&$\sim1$\\
			\multicolumn{2}{l}{Density ratio, $\rho_{d}/\rho_{a}$}&$0.1$-$100$&$\sim0.1$-$10$&$\sim1$\\
			\multicolumn{2}{l}{Mag.Reynolds, $Rm$}&$\sim10^{6}$&$\gg1$&$\gg1$\\
			\multicolumn{2}{l}{Peclet, $Pe$}&$\sim100$&$\gg1$&$\gg1$\\
			\multicolumn{2}{l}{Cooling, $\tau_{r}/\tau$ }&$\sim3000$&$<1$&$<1$\\
		\end{tabular}
	\end{ruledtabular}
\end{table}

The necessity and effectiveness of this laboratory scaling are to be discussed. Necessity comes from the fact that star ambient is indeed a time-varying, coupled physics system: for example, comparing very young YSO in cluster Serpens South\cite{plunkett2015} with the more evolved mature YSO HH34\cite{cabrit2000}, the visible knotty jet structure and position-velocity shape vary with age, which is related to the expelling of background cloud; PN transition from round to elliptical and to butterfly shaped shell can be attributed to different initial state of its host envelope, with portion of magnetic field's contribution to the evolutionary dynamics in debate\cite{soker2006}. A comprehensive consideration of magnetic field, drive strength, and background density is necessary. On the other hand, effectiveness is guaranteed by a set of dimensionless parameters\cite{ryutov1999,ryutov2000} listed in Table.~\ref{tab:table1}. Peclet $P_{e}$ and magnetic Reynolds number $R_{m}$ are much greater than unity, indicating all systems are convection dominated. Laboratory dimensionless parameters related to morphological changes, such as Alfven Mach numbers or density ratio, can effectively cover its corresponding range in YSO and PN even with multiple numbers combined.

To summarize, this article presents a laboratory magnetized plasma platform on which large parameter space scan of HD-MHD hybrid dynamics is performed. Results verify the robustness of stellar magnetic field collimation hypothesis under high background density, showing the ability to fit various observations under a common physical framework by assuming possible flow-field fine scale structures.

\bibliography{MHDjet_201905}

\end{document}